\documentclass[12pt, twoside]{article}
\pdfoutput=1
\usepackage{a4wide,amssymb,cite}
\usepackage{epsfig,graphicx}
\usepackage[usenames,dvipsnames]{color}
\usepackage{slashed}

\newcommand{\be}{\begin{equation}}
\newcommand{\ee}{\end{equation}}
\newcommand{\bea}{\begin{eqnarray}}
\newcommand{\eea}{\end{eqnarray}}

\begin{document}

\color{black}

\begin{flushright}
KIAS-P13006
\end{flushright}

\vspace{1cm}
\begin{center}
{\huge\bf\color{black}  Running inflation with unitary Higgs }\\
\bigskip\color{black}\vspace{1.5cm}{
{\large\bf Hyun Min Lee}
\vspace{0.5cm}
} \\[7mm]

{\it School of Physics, KIAS, Seoul 130-722, Korea.  }\\
\end{center}
\bigskip
\centerline{\large\bf Abstract}
\begin{quote}\large
We consider the renormalization group(RG) improved inflaton potential in unitarized Higgs inflation where the original Higgs inflation is unitarized by the addition of a real singlet scalar of sigma-model type.
The sigma field coupling to the Higgs, which is introduced to reproduce a large non-minimal coupling of the Higgs below the sigma scalar threshold, also improves the Standard Model vacuum stability due to the RG running.
Furthermore, the same sigma field coupling determines the reheating temperature or the number of efoldings.
Considering the uncertainties in the number of efoldings in the model, we show that the loop-corrected spectral index and tensor-to-scalar ratio are consistent with nine-year WMAP and new Planck data within $1\sigma$.

\end{quote} 

\thispagestyle{empty}

\normalsize

\newpage

\setcounter{page}{1}

\section{Introduction}

Theoretical problems in the standard Big Bang cosmology such as horizon, homogeneity, flatness and relics have motivated to introduce the early period of cosmic inflation \cite{inflation} by the addition of a scalar field, the so called inflaton, for the vacuum energy to dominate the universe. The initial condition for the large scale structure is set by the quantum fluctuation of the inflaton during inflation so the resultant post-inflation cosmology is then  well described. The recent nine-year WMAP data \cite{WMAP9} shows the evidence for a tilt in the primordial spectrum at the $5\sigma$ level, constraining the inflation models with more precision.  
Furthermore, the first results on the measurement of cosmological parameters by PLANCK \cite{Planck} have been released and have given strong constraints on inflation models \cite{Planckdata1,Planckdata2}. Thus, we are entering the era of precision cosmology to probe the remnants of the cosmic inflation, as has been the case in the Standard Model(SM) of particle physics.

Higgs inflation \cite{higgsinflation} has drawn much attention from both particle physics and cosmology communities, as a Higgs-like boson has been discovered recently at the Large Hadron Collider \cite{higgsdiscovery} and the model links between the SM and the inflation period in a minimal way. The Higgs doublet has a large non-minimal coupling to gravity \cite{non minimal} so the Higgs boson plays a role of the inflaton at large field values. However, there is a drawback in the original Higgs inflation due to the unitarity problem \cite{unitarity1}.
Although the unitarity cutoff during inflation is larger than the one in the vacuum \cite{unitarity2},  an inflationary plateau beyond the unitarity cutoff is still not justified under the perturbative expansion \cite{gianlee,Sthreshold}. Therefore, an extra degree of freedom at the unitarity scale is introduced to restore the unitarity \cite{gianlee} or an appropriate counterterm without extra degrees of freedom is required to cancel the dangerous interactions coming from the non-minimal coupling \cite{jmcdonald}.

As inflation occurs at the high energy scales, it is important to perform the renormalization group(RG) running from low energy until the inflation scale, to compare with the low energy parameters \cite{RG1,RG2}.
In particular, the stability of the SM vacuum requires the Higgs quartic coupling to remain positive at high scales \cite{vacuum stability,vs2}. In the original Higgs inflation, the standard RG analysis is trustable only up to the unitarity cutoff in the vacuum while the inflation occurs at the Higgs field values higher than the unitarity cutoff.
Although the effect of the RG running between the unitarity cutoff and the inflation scale has been assumed to be small, the unknown dynamics restoring the unitarity below the unitarity cutoff could spoil the inflationary plateau and/or the inflationary predictions.

In this paper, we consider the unitarized Higgs inflation proposed in Ref.~\cite{gianlee}, where a singlet scalar field of sigma-model type is introduced to restore the unitarity.  
In this model, unitarity is preserved all the way to the Planck scale and gives rise to the effective action of the Higgs inflation below the sigma scalar threshold. In the full theory, even if the Higgs doublet has a non-minimal coupling of order one, a large non-minimal coupling of the sigma field makes inflation
possible along a new flat direction, which is a linear combination of the Higgs and sigma fields.
Since the sigma field values for inflation are smaller than the Planck scale, the perturbative expansion is believed to be valid above the sigma scalar threshold too. 
We obtain the one-loop RG improved inflaton potential with the RG equations modified in the presence of a large non-minimal coupling for the sigma field. The vacuum stability can be improved by the RG running due to the mixing coupling between the Higgs and sigma fields, provided that the SM vacuum is guaranteed to be stable at the sigma scalar threshold. 
There are uncertainties in the number of efoldings because the reheating temperature depends on the sigma coupling to the Higgs. Taking this result into account, we show that the loop-corrected spectral index, controlled by the same sigma coupling, and the tensor-to-scalar ratio, are consistent with the nine-year WMAP data within $1\sigma$.

The paper is organized as follows. We first begin with the description of the unitarized Higgs inflation and discuss the effective theories at low energy and during inflation. Then, we compute the one-loop Coleman-Weinberg corrections for the RG-improved inflaton potential and show how the vacuum stability is improved in this model. Next we present the results of the spectral index,  the tensor-to-scalar ratio, the running of the spectral index, etc.  In the next section, we give a brief discussion on the reheating temperature in relation to the number of efoldings. Finally conclusions are drawn. There is one appendix containing the RG equations applicable to the energy scales above the sigma field threshold.

\section{Inflation with non-minimal coupling and unitary Higgs}

In order to solve the unitarity problem, we require extra dynamical degrees of freedom to restore unitarity without ruining the flat plateau. A UV complete model with a real singlet scalar of sigma-model type was proposed in~\cite{gianlee} and it has been shown that extra singlet coupling could also solve simultaneously the vacuum instability problem in the SM \cite{Sthreshold,lebedev}.

The Jordan-frame Lagrangian of the model is
\bea
\frac{{\cal L}_J}{\sqrt{-g_J}}&=&\frac{1}{2} \Big(M^2+\xi\sigma^2+2\zeta  H^\dagger H\Big)R-\frac{1}{2}(\partial_\mu\sigma)^2-|D_\mu H|^2 \nonumber \\
&&-\frac{1}{4}\lambda_\sigma \Big(\sigma^2- \omega^2+2\frac{\lambda_{H\sigma}}{\lambda_\sigma} H^\dagger H\Big)^2-\Big(\lambda_H-\frac{\lambda^2_{H\sigma}}{\lambda_\sigma}\Big) \Big(H^\dagger H-\frac{v^2}{2}\Big)^2 \ ,
\label{jordanaction0}
\eea
where $M,\omega$ and $v$ are mass parameters with $v\ll M,\omega$ (so that the $\sigma$ field is heavy)
and $\xi,\zeta$ are positive non-minimal couplings with $\xi\gg \zeta$. 

The large nonzero vev of $\sigma$, $\langle\sigma\rangle\simeq \omega$, is crucial to make the unitarity cutoff $\Lambda_{UV}$ larger. It is straightforward to find that 
\be
\Lambda_{UV}=  \Big(1+6r \xi\Big)\frac{M_{\rm Pl}}{\xi}\ ,
\label{UVcutoff}
\ee
where the Planck mass is now $M_{\rm Pl}^2= M^2+\xi\omega^2$, and we measure the contribution of the $\sigma$ vev by the ratio $r= \xi \omega^2/M_{\rm Pl}^2$, which in general can take values from 0 to 1. One can see how the cutoff is pushed up to $r M_{\rm Pl}$ for moderate values of $r\gtrsim 1/\xi$.

In the following discussion,  for simplicity, we consider a simplified verison of the unitarized Higgs inflation where the tree-level Einstein term and the non-minimal coupling for the Higgs doublet is absent, $M=0$ and $\zeta=0$, in Jordan frame.
Then, the Jordan-frame action in unitary gauge with $H=(0,\phi)^T/\sqrt{2}$ is 
\bea
\frac{{\cal L}_J}{\sqrt{-g_J}}=\frac{1}{2}\xi\sigma^2 R-\frac{1}{2}(\partial_\mu \sigma)^2-\frac{1}{2}(\partial_\mu \phi)^2 -V_J  \label{jordanaction}
\eea
where 
\be
V_J=\frac{1}{4}\lambda_\sigma \Big(\sigma^2-\omega^2+\frac{\lambda_{H\sigma}}{\lambda_\sigma} \phi^2\Big)^2+\frac{1}{4}\Big(\lambda_H-\frac{\lambda^2_{H\sigma}}{\lambda_\sigma}\Big)(\phi^2-v^2)^2
\ee
and $\omega\equiv \frac{M_P}{\sqrt{\xi}}$ is chosen to reproduce the Jordan-frame action of the Higgs inflation with a positive non-minimal coupling $\xi_h=-\frac{\lambda_{H\sigma}}{\lambda_\sigma}\,\xi$ for $\lambda_{H\sigma}<0$, after integrating out the $\sigma$ field by $\sigma^2=-\frac{\lambda_{H\sigma}}{\lambda_\sigma} \phi^2+\omega^2$. The mass of the $\sigma$ field in the vacuum is given by
\be
M_{\bar\sigma}^2=\lambda_\sigma \frac{2 r M_{\rm Pl}^2}{(1+6r\xi)\xi}\simeq\lambda_\sigma\frac{M_{\rm Pl}^2}{3\xi^2}
\ee
where $\bar\sigma$ denotes the canonically normalized field. The COBE constraint  precisely fixes the sigma mass in the vacuum to be $M_{\bar\sigma}\approx 10^{13}\,{\rm GeV}$ \cite{Sthreshold}.

Below the sigma mass scale, the effective action in Jordan frame is
\be
\frac{{\cal L}_J}{\sqrt{-g_J}}=\frac{1}{2}\Big(M^2_P+\xi_{\rm eff}\phi^2\Big) R-\frac{1}{2}(\partial_\mu \phi)^2
-\frac{1}{4}\lambda_{\rm eff} (\phi^2-v^2)^2
\ee
where
the effective non-minimal coupling $\xi_{\rm eff}$ and quartic coupling $\lambda_{\rm eff}$ for the Higgs  are matched to the fundamental couplings as
\bea
\xi_{\rm eff}&\equiv&- \frac{\lambda_{H\sigma}}{\lambda_\sigma}\,\xi,   \label{xieff}\\
 \lambda_{\rm eff}&\equiv& \lambda_H -\frac{\lambda^2_{H\sigma}}{\lambda_\sigma}. \label{treeshift}
\eea
Thus, the Higgs quartic coupling $\lambda_H$ can be larger than the SM value inferred from the Higgs mass, helping to ensure the vacuum stability at large field values when the sigma field is lighter than the instability scale $\Lambda_I$ in the SM \cite{Sthreshold}.  Thus, we are forced to the Higgs masses for which $\Lambda_I> 10^{13}$ GeV. This requires $m_h>125$ GeV (at 90\% CL in $M_t$ from the kinematical top mass at the Tevatron and $\alpha_{\rm s}$) \cite{Sthreshold}, which is marginally compatible with the Higgs-like boson discovered by ATLAS and CMS \cite{higgsdiscovery}. But, we note that the vacuum stability bound on the Higgs mass depends on the top pole mass, which still has a large uncertainty as suggested from the top pair production cross section measurements at the Tevatron\cite{djouadi,masina}. 
In this work, we assume that the SM vacuum is stable within the uncertainties of the top pole mass at the sigma field threshold and consider the possibility that the loop corrections of the sigma field help improve the vacuum stability.

Performing a Weyl scaling of the metric, we obtain the Einstein-frame action from eq.~(\ref{jordanaction}) as follows,
\bea
\frac{{\cal L}_E}{\sqrt{-g_E}}=\frac{1}{2}M^2_P R-\frac{1}{2}\Big(\frac{\omega}{\sigma}\Big)^2\bigg[(1+6\xi)(\partial_\mu\sigma)^2+(\partial_\mu\phi)^2\bigg]-\frac{1}{4}\Big(\frac{\omega}{\sigma}\Big)^4 V_J.
\eea 
Redefining the fields by $\sigma\equiv \omega\, e^{\chi/\sqrt{6}M_P}$ and ${\tilde\phi}\equiv\omega\phi/\sigma=\phi\,e^{-\chi/\sqrt{6}M_P}$, the above action becomes
\bea
\frac{{\cal L}_E}{\sqrt{-g_E}}=\frac{1}{2}M^2_P R-\frac{1}{2}\Big(1+\frac{1}{6\xi}+\frac{{\tilde\phi}^2}{6M^2_P}\Big)(\partial_\mu\chi)^2-\frac{1}{2}(\partial_\mu{\tilde\phi})^2-\frac{1}{\sqrt{6}}\,\frac{{\tilde\phi}}{M_P}\,\partial_\mu\chi\partial^\mu {\tilde\phi} -V_E \label{einaction}
\eea
with
\be
V_E=\frac{1}{4}\omega^4\lambda_\sigma\bigg(1-e^{-2\chi/\sqrt{6}M_P}+\frac{\lambda_{H\sigma}}{\lambda_\sigma}\frac{{\tilde\phi}^2}{\omega^2}\bigg)^2+\frac{1}{4}\Big(\lambda_H-\frac{\lambda_{H\sigma}^2}{\lambda_\sigma}\Big)\bigg(
{\tilde\phi}^2-v^2\,e^{-2\chi/\sqrt{6}M_P}\bigg)^2. \label{einpot}
\ee
We note that the kinetic terms for the sigma and Higgs fields is of sigma-model type, with the coset space described by $SO(1,5)/SO(5)$. 

Taking $|\sigma|\gg \omega$, the Einstein-frame potential approximates the potential for $\tilde\phi$,  
\be
V_E\simeq \frac{1}{4}\bigg(\lambda_\sigma \omega^4+2\lambda_{H\sigma} \omega^2{\tilde \phi}^2+\lambda_H{\tilde\phi}^4\bigg).
\ee
Thus, for $\lambda_{H\sigma}<0$, the potential has two minima at ${\tilde\phi}=\pm\sqrt{-\frac{\lambda_{H\sigma}}{\lambda_H}}\,\omega\equiv\pm {\tilde\phi}_0$. Therefore, after stabilizing $\tilde\phi$ at one of the minima, we obtain the flat potential for $\chi$ as 
\be
V_E= V_0\Big(1-e^{-2\chi/\sqrt{6}M_P}\Big)^2,\quad V_0\equiv \frac{\omega^4}{4}\,\Big(\lambda_\sigma-\frac{\lambda^2_{H\sigma}}{\lambda_H}\Big).
\ee
Therefore, the sigma field drives a slow-roll inflation while the Higgs field is stabilized at a large VEV during inflation. The difference from a single-field inflation with non-minimal coupling is that the Higgs field contributes a large vacuum energy during inflation and participates in the reheating process as will be discussed in the later section.

Here, we note that a positive vacuum energy during inflation is obtained for $\lambda_H> \frac{\lambda^2_{H\sigma}}{\lambda_\sigma}$. Thus, the vacuum stability condition becomes the condition for the positive inflaton vacuum energy so it has not been improved at tree level, as compared to the SM, where the corresponding condition from the matching scale at the sigma mass scale is $\lambda_{\rm eff}=\lambda_H- \frac{\lambda^2_{H\sigma}}{\lambda_\sigma}>0$.
Since $|{\tilde\phi}_0|\ll M_P$ for $\xi\gg 1$,  the kinetic mixing term is ignored and both $\chi$ and $\tilde\phi$ are canonical scalar fields.
Heneceforth we set $M_P=1$.

\section{Effective inflaton potential}

We consider the one-loop Coleman-Weinberg corrections in unitarized Higgs inflation for the effective potential for inflation and discuss the effect of the sigma-field couplings on the vacuum stability.

\subsection{One-loop inflaton potential}

First, ignoring the contribution coming from the inflaton, we get  the one-loop Coleman-Weinberg potential as
\bea
V_{CW}&=&\frac{m^4_{\tilde\phi}}{64\pi^2}\Big(\ln\frac{m^2_{\tilde\phi}}{\mu^2}-\frac{3}{2}\Big)+\frac{3m^4_G}{64\pi^2}\Big(\ln\frac{m^2_G}{\mu^2}-\frac{3}{2}\Big)+\frac{6m^4_W}{64\pi^2}\Big(\ln\frac{m^2_W}{\mu^2}-\frac{5}{6}\Big) \nonumber \\
&&+ \frac{3m^4_Z}{64\pi^2}\Big(\ln\frac{m^2_Z}{\mu^2}-\frac{5}{6}\Big)
-\frac{3m^4_t}{16\pi^2}\Big(\ln\frac{m^2_t}{\mu^2}-\frac{3}{2}\Big) \label{CW}
\eea
where the ``effective'' masses for the heavy mode $\tilde\phi$, Goldstone bosons, W and Z bosons and top quark are given in order as
\bea
m^2_{\tilde\phi}&=& 3\lambda_H{\tilde\phi}^2+\lambda_{H\sigma}\omega^2\Big(1-e^{-2\chi/\sqrt{6}}\Big), \\
m^2_G&=&\lambda_H {\tilde\phi}^2 +\lambda_{H\sigma}\omega^2\Big(1-e^{-2\chi/\sqrt{6}}\Big), \\
m^2_W&=&\frac{1}{4}g^2\phi^2\Big(\frac{\omega}{\sigma}\Big)^2=\frac{1}{4}g^2{\tilde\phi}^2, \\
m^2_Z&=&\frac{1}{4}(g^2+g^{\prime 2})\phi^2\Big(\frac{\omega}{\sigma}\Big)^2=\frac{1}{4}(g^2+g^{\prime 2}){\tilde\phi}^2,\\
m^2_t&=&\frac{1}{2}y_t^2\phi^2\Big(\frac{\omega}{\sigma}\Big)^2=\frac{1}{2}y^2_t {\tilde\phi}^2.
\eea

From eq.~(\ref{CW}), we find that all logarithms contain the corrections to the effective quartic couplings for $\tilde\phi$ while only the Higgs portal term gives rise to the correction to the ``physical'' mass of $\tilde\phi$ evaluated at ${\tilde\phi}=0$ as follows,
\be
M^2_{\tilde\phi}=M^2_{\rm tree}+M^2_{\rm loop}
\ee
with
\bea
M^2_{\rm tree}&=&\lambda_{H\sigma}\omega^2\Big(1-e^{-2\chi/\sqrt{6}}\Big),\\
M^2_{\rm loop}&=&\frac{3}{16\pi^2}\lambda_H\lambda_{H\sigma}\omega^2\Big(1-e^{-2\chi/\sqrt{6}}\Big)\ln\frac{m^2_{\tilde\phi}}{\mu^2}.
\eea
We note that the loop mass has the same functional form for $\chi$ as for the tree-level mass and it can be absorbed by renormalizing the tree-level mass parameter. In the end of inflation, the inflaton rolls down to the minimum of the potential at $\chi\simeq 0$, so the large mass terms for $\tilde\phi$ vanish.

Using the equation of motion for $\tilde\phi$, 
\be
{\tilde\phi}^2=-\frac{\lambda_{H\sigma}}{\lambda_H}\omega^2\Big(1-e^{-2\chi/\sqrt{6}}\Big),  \label{eom}
\ee
the Goldstone boson masses vanish while
the effective masses of the rest become
\bea
m^2_{\tilde\phi}&=&-2\lambda_{H\sigma}\omega^2\Big(1-e^{-2\chi/\sqrt{6}}\Big),\\
m^2_W&=& -\frac{1}{4}g^2\,\frac{\lambda_{H\sigma}}{\lambda_H}\omega^2 \Big(1-e^{-2\chi/\sqrt{6}}\Big), \\
m^2_Z&=&-\frac{1}{4}(g^2+g^{\prime 2})\,\frac{\lambda_{H\sigma}}{\lambda_H}\omega^2 \Big(1-e^{-2\chi/\sqrt{6}}\Big), \\
m^2_t&=&-\frac{1}{2}y^2_t \,\frac{\lambda_{H\sigma}}{\lambda_H}\omega^2 \Big(1-e^{-2\chi/\sqrt{6}}\Big).
\eea
Here, we note that all the masses are of the same form as in the SM without non-minimal coupling but with the Higgs being replaced by ${\tilde\phi}(\chi)$.
After plugging the above masses in eq.~(\ref{CW}), we obtain the one-loop corrected inflaton potential  renormalized at $\mu={\tilde \phi}(\chi)$ with eq.~(\ref{eom}) as follows,
\bea
V(\chi)={\hat V}_0 (1-e^{-2\chi/\sqrt{6}})^2  \label{effpot}
\eea
where the effective vacuum energy during inflation ${\hat V}_0$ is given by
\bea
{\hat V}_0&=&\frac{\omega^4}{4 }\bigg[\Big(\lambda_\sigma-\frac{\lambda^2_{H\sigma}}{\lambda_H}\Big)+\frac{1}{4\pi^2}\lambda^2_{H\sigma}\Big(\ln(2\lambda_H)-\frac{3}{2}\Big)+\frac{1}{16\pi^2}\frac{\lambda^2_{H\sigma}}{\lambda^2_H}
\bigg\{\frac{3}{8}g^4 \Big(\ln\Big(\frac{g^2}{4}\Big) -\frac{5}{6}\Big)\nonumber \\
&&\quad+\frac{3}{16}(g^2+g^{\prime 2})^2\Big( \ln\Big(\frac{1}{4}(g^2+g^{\prime 2})\Big)-\frac{5}{6}\Big)-3 y^4_t \Big(\ln\Big(\frac{y^2_t}{2}\Big)-\frac{3}{2}\Big) \bigg\}\bigg]  \label{effenergy}
\eea
where all the running couplings are evaluated at $\mu={\tilde\phi}$. 
Consequently, the effective inflaton potential is determined by the Higgs quartic coupling $\lambda_H$, the extra quartic couplings, $\lambda_\sigma$, $\lambda_{H\sigma}$, the SM gauge couplings and the top Yukawa coupling.

\subsection{Sigma-field coupling and vacuum stability}

\begin{figure}[t]
\centering
\includegraphics[width=8cm]{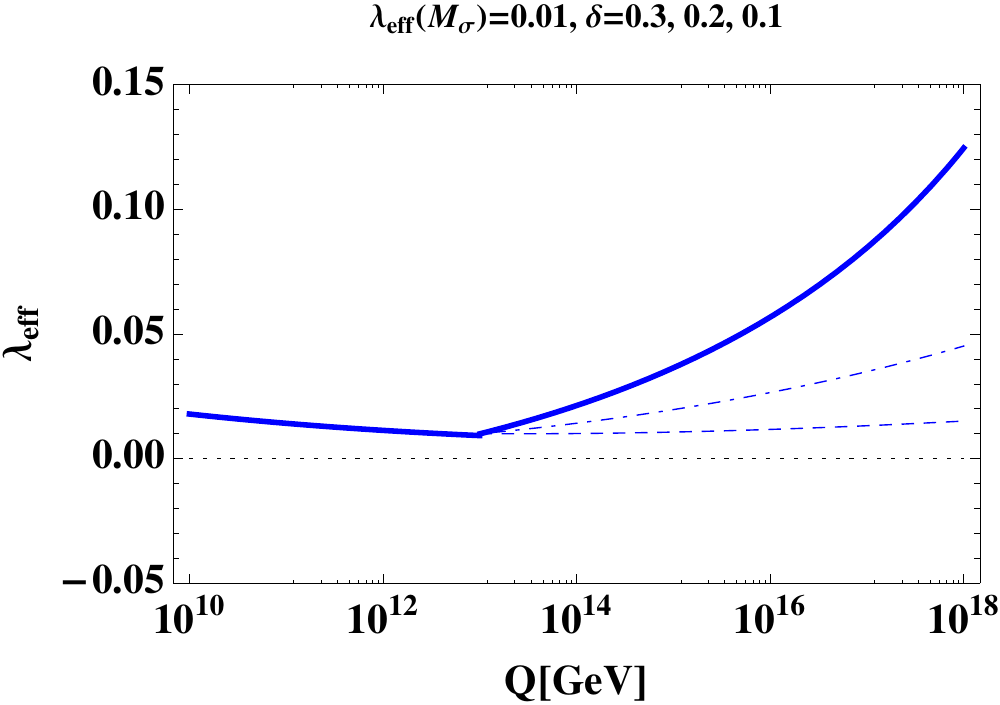}
\includegraphics[width=7.8cm]{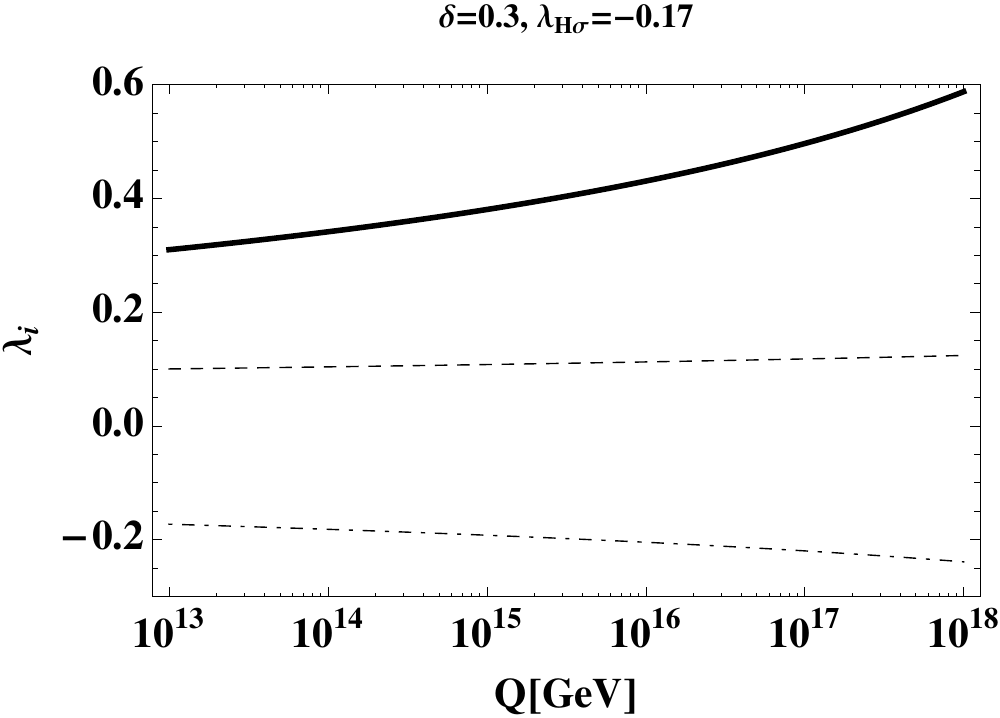}
\caption{Left: RG running of the effective Higgs quartic coupling, $\lambda_{\rm eff}=\lambda_H-\lambda^2_{H\sigma}/\lambda_\sigma$. The Higgs quartic coupling $\lambda_H$ is matched to the effective Higgs quartic coupling $\lambda_{\rm eff}=0.01$ at sigma field threshold of $M_{\sigma}=10^{13}\,{\rm GeV}$. $\delta\equiv \lambda^2_{H\sigma}/\lambda_\sigma$ is taken to $0.3, 0.2, 0.1$ from top to bottom  for fixed $\lambda_{H\sigma}=-0.17$.
Right: RG running of the quartic couplings, $\lambda_H, \lambda_\sigma$ and $\lambda_{H\sigma}$, from top to bottom, for $\delta=0.3$ and $\lambda_{H\sigma}=-0.17$.
}
\label{fig:RG}
\end{figure}

As discussed before, positivity of the tree-level inflaton potential requires $\lambda_H>\lambda^2_{H\sigma}/\lambda_\sigma$, which is the same as the vacuum stability condition in the SM, $\lambda_{\rm eff}>0$, from eq.~(\ref{treeshift}).
However, the effect of the running quartic couplings and the threshold corrections to the effective vacuum energy may make the vacuum energy larger. Thus, the vacuum stability can be guaranteed during inflation, once ensured at the sigma mass scale.

From appendix A, the RG equation for $\lambda_{\rm eff}\equiv \lambda_H-\lambda^2_{H\sigma}/\lambda_\sigma$ with $\delta\lambda\equiv \lambda^2_{H\sigma}/\lambda_\sigma$ is
\bea
\frac{d\lambda_{\rm eff} }{d\ln\mu} \approx \beta^{\rm SM}_{\lambda_{\rm eff}}
+\frac{8}{(4\pi)^2}(3\lambda_{\rm eff}+\delta\lambda)\,\delta\lambda\,.
\eea
Thus, due to the positive contribution coming from the sigma-field couplings in the RG equation, the vacuum instability scale gets higher. On the left of Fig.~\ref{fig:RG}, we depict the running of the effective Higgs quartic coupling above the sigma field threshold at $M_\sigma=10^{13}\,{\rm GeV}$, depending on the tree-level shift in the effective Higgs quartic coupling. We also show on the right of Fig.~\ref{fig:RG} that the quartic couplings producing a sizable shift $\delta=0.3$ in the effective Higgs quartic coupling remain perturbative all the way to the Planck scale.

\section{Corrections to the inflationary observables}

In the one-loop improved inflaton potential, the effective vacuum energy has threshold corrections coming from the heavy modes of non-inflaton fields coupled to the inflaton field. Since the threshold corrections depend on the running couplings, in turn, the inflaton field value, they can give extra contributions to the spectral index.
By using the RG equations given in appendix A, we consider the threshold corrections to the spectral index and other inflationary observables in this section. Furthermore, we discuss the reheating temperature and the predicted number of efoldings in the model, depending on the mixing coupling between the Higgs and sigma fields.

\subsection{Spectral index and tensor-to-scalar ratio}

The slow-roll parameters are 
\bea
\epsilon=\frac{M^2_P}{2} \Big(\frac{V'}{V}\Big)^2, \quad \eta=M^2_P\, \frac{V^{\prime\prime}}{V}.
\eea
The slow-roll conditions are $\epsilon\ll 1$ and $|\eta|\ll 1$. The first condition(``slowly varying") corresponds to making the Hubble parameter during inflation proximate to constant while the second condition comes from the slowly varying condition plus $3H{\dot\chi}=-V'$(``slow-roll approximations"). Then, the spectral index and the tensor-to-scalar ratio are then evaluated at the horizon exit, according to the following,
\be
n_s=1-6\epsilon+2\eta, \quad\quad   r=16 \epsilon.
\ee

\begin{figure}[t]
\centering
\includegraphics[width=8cm]{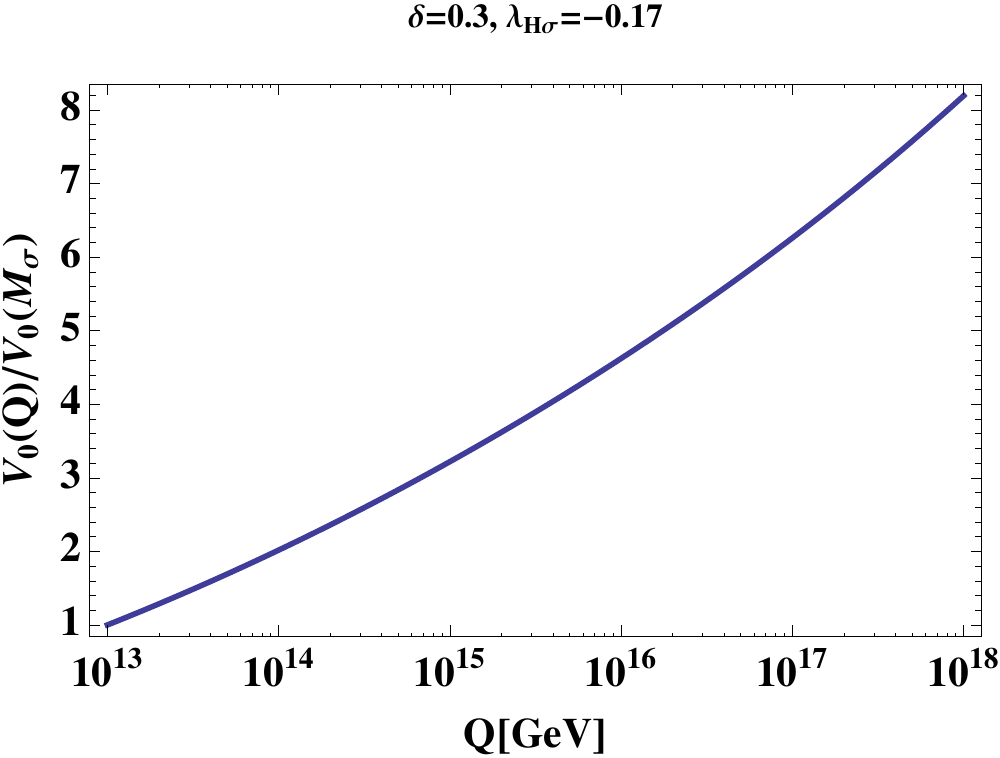}
\caption{RG running of the inflaton vacuum energy. We have fixed $\delta=\lambda^2_{H\sigma}/\lambda_\sigma=0.3$ and $\lambda_{H\sigma}=-0.17$.
}
\label{fig:V0run}
\end{figure}

When we compute the slow-roll parameters, we take the tree-level vacuum energy to be dominant over the Coleman-Weinberg correction. Then, the inflaton potential is given by the tree-level one with the running couplings:
\be
V(\chi)= {\hat V}_0 (1-e^{-2\chi/\sqrt{6}})^2, \quad {\hat V}_0\approx \frac{1}{4\xi^2(\mu(\chi))}\Big(\lambda_\sigma(\mu(\chi)) -\frac{\lambda^2_{H\sigma}(\mu(\chi))}{\lambda_H(\mu(\chi))}\Big) 
\ee
where the couplings depend on the canonical inflaton field $\chi$ through the renormalization condition, $\mu={\tilde\phi}(\chi)$ with eq.~(\ref{eom}). 
In Fig.~\ref{fig:V0run}, we show the RG scale dependence of the inflaton vacuum energy. We note that from the renormalization condition $\mu={\tilde\phi}$ with eq.~(\ref{eom}), the RG scale during inflation is of order $|\lambda_{H\sigma}|M_P /(\lambda_H \sqrt{\xi})$, which is about $10^{16}\,{\rm GeV}$ for $\lambda_H\sim \lambda_{H\sigma}$ from the COBE normalization as will be discussed later in this section.

Then, the field  derivatives of the potential are
\bea
\frac{dV}{d\chi}&=& \Big(\frac{\partial {\hat V}_0}{\partial \ln\mu} +4 {\hat V}_0\Big) \,\frac{1}{\sqrt{6}} e^{-2\chi/\sqrt{6}} (1-e^{-2\chi/\sqrt{6}}  ), \\
\frac{d^2V}{d\chi^2}&=&-  \Big(\frac{\partial {\hat V}_0}{\partial \ln\mu} +4 {\hat V}_0\Big)\,\frac{1}{3} e^{-2\chi/\sqrt{6}} (1-2 e^{-2\chi/\sqrt{6}}  ) \nonumber \\
&&\quad+ \frac{\partial}{\partial\ln \mu}  \Big(\frac{\partial {\hat V}_0}{\partial \ln\mu} +4 {\hat V}_0\Big)\, \frac{1}{6} e^{-4\chi/\sqrt{6}}
\eea
where use is made of $\partial \ln \mu/\partial\chi=\frac{1}{\sqrt{6}} e^{-2\chi/\sqrt{6}}/(1-e^{-2\chi/\sqrt{6}})$ for $\mu={\tilde\phi}$ in the chain rule for $\frac{\partial {\hat V}_0}{\partial \chi}=\frac{\partial {\hat V}_0}{\partial{\ln \mu}}\frac{\partial \ln \mu}{\partial \chi}$.
Therefore, we get the slow-roll parameters as
\bea
\epsilon&=& \Big(1+\frac{1}{4{\hat V}_0}\frac{\partial {\hat V}_0}{\partial \ln\mu}\Big)^2 \,\frac{4}{3} e^{-4\chi/\sqrt{6}}, \\
\eta&=&(1-e^{-2\chi/\sqrt{6}})^{-2}\bigg[-\Big(1+\frac{1}{4{\hat V}_0}\frac{\partial {\hat V}_0}{\partial \ln\mu}\Big)\, \frac{4}{3}e^{-2\chi/\sqrt{6}}\nonumber \\
&&\quad+\frac{4}{3}\Big(2+\frac{1}{{\hat V}_0}\frac{\partial {\hat V}_0}{\partial \ln\mu}\Big) \,e^{-4\chi/\sqrt{6}}+\frac{1}{6{\hat V}_0}\frac{\partial ^2{\hat V}_0}{\partial (\ln\mu)^2}\, e^{-4\chi/\sqrt{6}}\bigg]
\eea
where use is made of the RG equations in appendix A to get
\bea
\frac{\partial {\hat V}_0}{\partial \ln\mu}&=&\sum_i \beta_i \frac{\partial {\hat V}_0}{\partial\lambda_i} \nonumber \\
&=&\frac{\lambda_{H\sigma}^2}{64\pi^2\xi^2}\bigg[ 8+\frac{1}{\lambda^2_H}\Big(\frac{3}{8}(2g^4+(g^{\prime 2}+g^2)^2 )-6 y^2_t\Big)\bigg].
\eea
So, the loop corrections to the slow-roll parameters are determined by $\delta\lambda$, $\lambda_H$ and the gauge and top Yukawa couplings. When we make use of the effective Higgs quartic coupling at the matching scale by $\lambda_{\rm eff}=\lambda_H-\delta\lambda$, there is only one unknown parameter, $\lambda_H$ or $\delta\lambda$.
Here, we note that the running effect of the non-minimal coupling $\xi$ is suppressed by $1/\xi$.
We also note that the second derivative of the vacuum energy is given by
\be
\frac{\partial^2 {\hat V}_0}{\partial (\ln\mu)^2}=\sum_i\Big(\beta'_i \frac{\partial {\hat V}_0}{\partial\lambda_i}+\beta_i\sum_j \beta_j \frac{\partial^2 {\hat V}_0}{\partial\lambda_i \partial\lambda_j}\Big)\sim \frac{\beta_i}{\lambda_i} \frac{\partial {\hat V}_0}{\partial\ln\mu}\ll \frac{\partial {\hat V}_0}{\partial\ln\mu} 
\ee
where $\beta_i/\lambda_i\ll 1$ is assumed in the last inequality. Thus, we ignore the second derivative terms with respect to $\ln\mu$ in the slow-roll parameters.

We can compute the total number of e-foldings as
\be
N=\int^{t_f}_{t_i}H dt=\int^{\chi_i}_{\chi_f} \frac{d\chi}{\sqrt{2\epsilon}}\simeq \frac{3}{4|A|}\Big(e^{2\chi_i/\sqrt{6}}-e^{2\chi_f/\sqrt{6}}\Big) \label{nefolding}
\ee
where
\bea
A&\equiv& 1+\frac{1}{4{\hat V}_0}\frac{\partial {\hat V}_0}{\partial \ln\mu} \nonumber \\
&\simeq&1+\frac{\delta\lambda}{16\pi^2}\frac{\lambda_H}{\lambda_H-\delta\lambda}\bigg[ 8+\frac{1}{\lambda^2_H}\Big(\frac{3}{8}(2g^4+(g^{\prime 2}+g^2)^2 )-6 y^2_t\Big)\bigg].
\eea
We note that the $\chi$ dependence of $A$ in the integrand can be ignored, because $\delta\lambda_i=\beta_i \,\delta\ln\mu\sim \beta_i \frac{1}{\sqrt{6}} e^{-2\chi/\sqrt{6}}\delta\chi$ is exponentially suppressed during inflation.
Here, $e^{2\chi_f/\sqrt{6}}=2 |A|/\sqrt{3}$ corresponds to the field value at which
 $\epsilon=1$ when the slow-roll dynamics ends. So, we need $|A|>\frac{\sqrt{3}}{2}$ from $e^{2\chi_f/\sqrt{6}}>1$. On the other hand, from eq.~(\ref{nefolding}), we also get $e^{2\chi_i/\sqrt{6}}=\frac{2}{3} |A|(2N+\sqrt{3})$ at the horizon exit. Since we can determine $\lambda_H$ in terms of $\lambda_{\rm eff}$ and $\delta\lambda$ at the sigma threshold and the top pole mass determines $y_t$ with some uncertainties, the loop corrections eventually depend on the unknown $\delta\lambda$ for a given $\lambda_{\rm eff}$ at the sigma field threshold.  The same parameter $\delta\lambda$ controls the vacuum stability above the sigma field threshold as discussed in section 3.2.

  \begin{figure}[t]
\centering
\includegraphics[width=7.5cm]{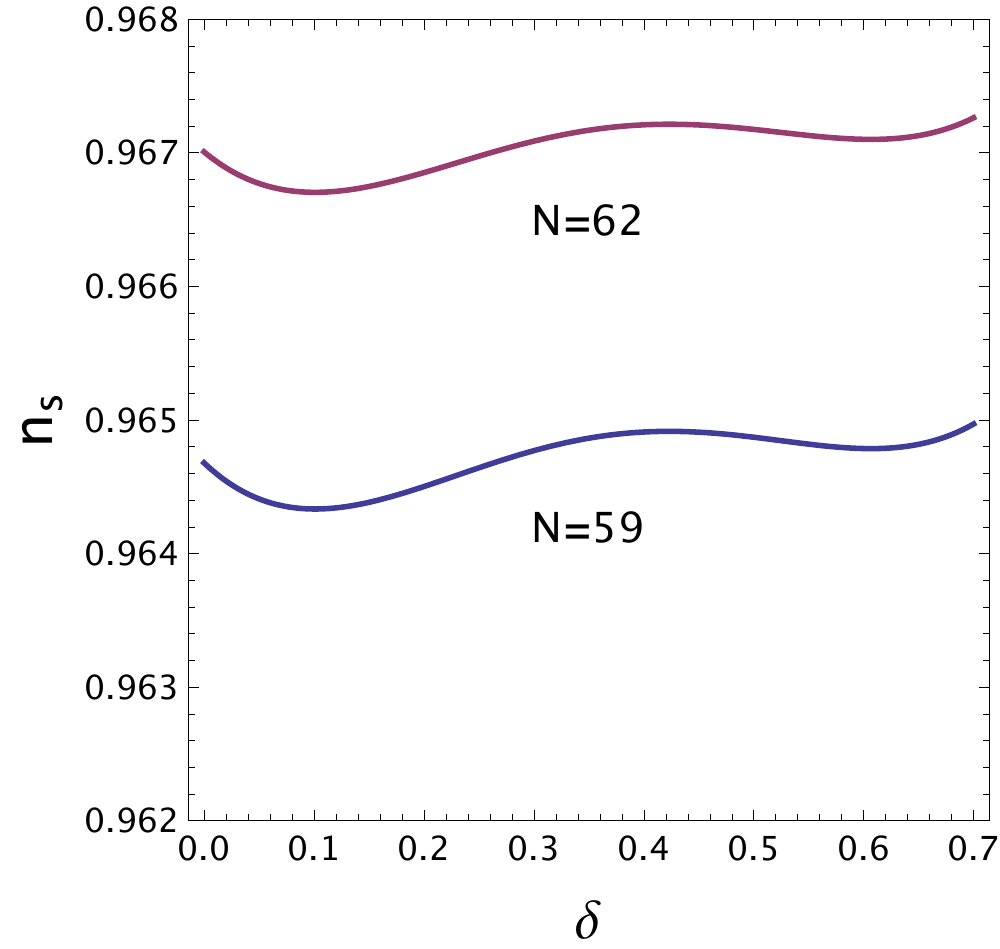}
\includegraphics[width=7.9cm]{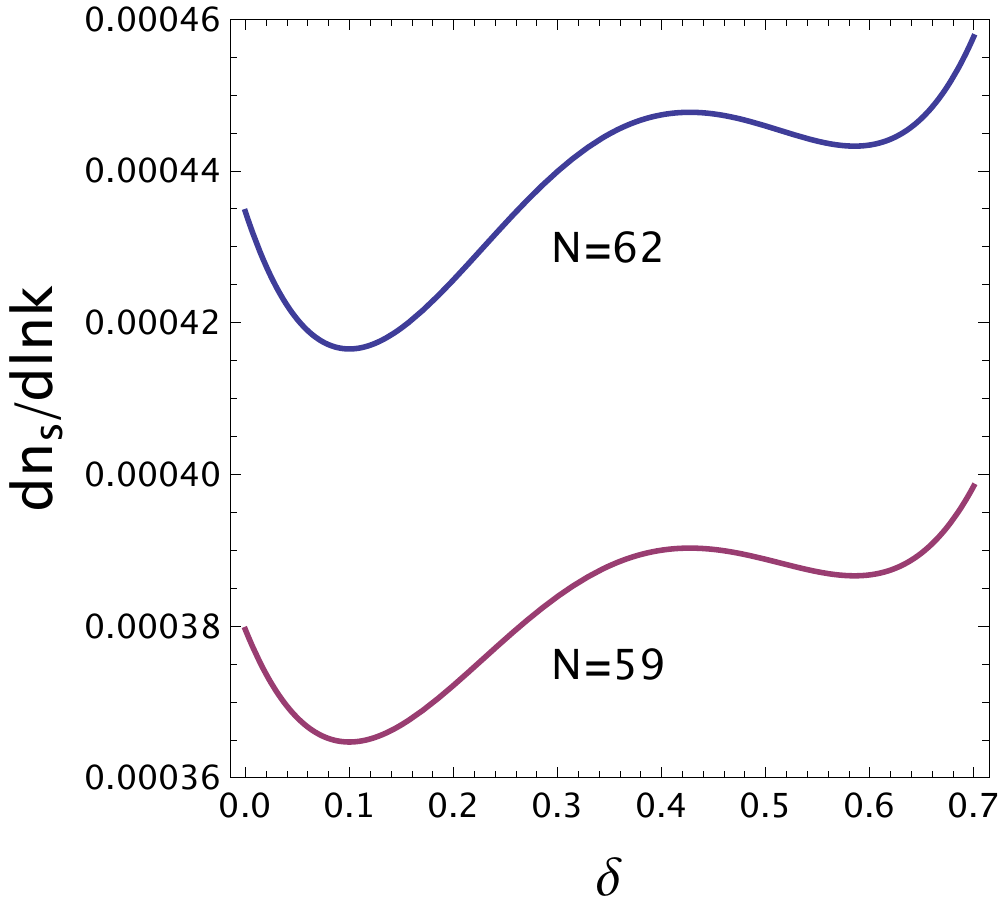}
\caption{Left: Spectral index $n_s$(Left) and running of spectral index(Right) vs the tree-level shift in the Higgs quartic coupling, $\delta\equiv\lambda^2_{H\sigma}/\lambda_\sigma$. 
The number of efoldings is taken to $N=59$ or $62$. We have set $\lambda_{\rm eff}(M_\sigma)=0.01$ and $\lambda_\sigma(M_\sigma)=0.1$ to evaluate the loop corrections in $A$ at the inflation scale, although the results do not depend on $\lambda_\sigma$ apart from perturbativity.
}
\label{fig:ns}
\end{figure}
 
Then, we can rewrite the slow-roll parameters in terms of the number of e-foldings, 
\bea
\epsilon&=&  \frac{3}{(2 N+\sqrt{3})^2},   \label{epsilon}\\
 \eta&=&\Big(1-\frac{3}{2|A|}\frac{1}{2N+\sqrt{3}}\Big)^{-2}\left( -\frac{2A}{|A|}\frac{1}{2N+\sqrt{3}}+\frac{6}{A^2}\frac{2A-1}{(2N+\sqrt{3})^2}\right). \label{eta}
\eea
Therefore, the spectral index becomes
\bea
n_s&\simeq& 1-\frac{2\Big(2A(2N+\sqrt{3})/|A|+9-6/A+6/A^2\Big)}{(2N+\sqrt{3})^2}.
\eea
For instance, for the tree-level potential with $\delta\lambda=0$, we obtain $n_s=0.9647-0.9670$ for $N=59-62$.
For comparison, the measured spectral index from nine-year WMAP with eCMB, BAO and $H_0$ is given by $n_s=0.9608\pm 0.0080$ \cite{WMAP9}. Furthermore, the recent Planck data combined with the WMAP large-angle polarization, indicates a more precise value of the spectral index 
as $n_s=0.9603\pm 0.0073$ \cite{Planckdata1}. Thus, the spectral index obtained in the model is consistent with Planck data, within $1\sigma$.
On the left of Fig.~\ref{fig:ns}, we show the dependence of the spectral index on the number of efoldings and the loop corrections as a function of the tree-level shift in the Higgs quartic coupling, $\delta\lambda$, and find that the obtained spectral index is consistent with the current observation within $1\sigma$.

In the unitarized Higgs inflation, the reheating temperature is sensitive to the Higgs component of the inflaton, which is determined by the mixing coupling $\lambda_{H\sigma}$ between the Higgs and sigma fields.  
The spectral index varies by $0.002$, depending on the number of efoldings, $N=59-62$, in our model, as will be discussed in the next section. We also note that for the number of efoldings being fixed, the loop corrections make $|\Delta n_s|$  less than $0.001$, so the effect of the loop corrections is much smaller than the one of the number of efoldings.
However, the expected sensitivity at PLANCK in measuring the spectral index is $\Delta n_s=\pm 0.004$ \cite{Planck} so it might not be possible to measure the reheating temperature or the loop corrections precisely by PLANCK yet.

\begin{figure}[t]
\centering
\includegraphics[width=7.5cm]{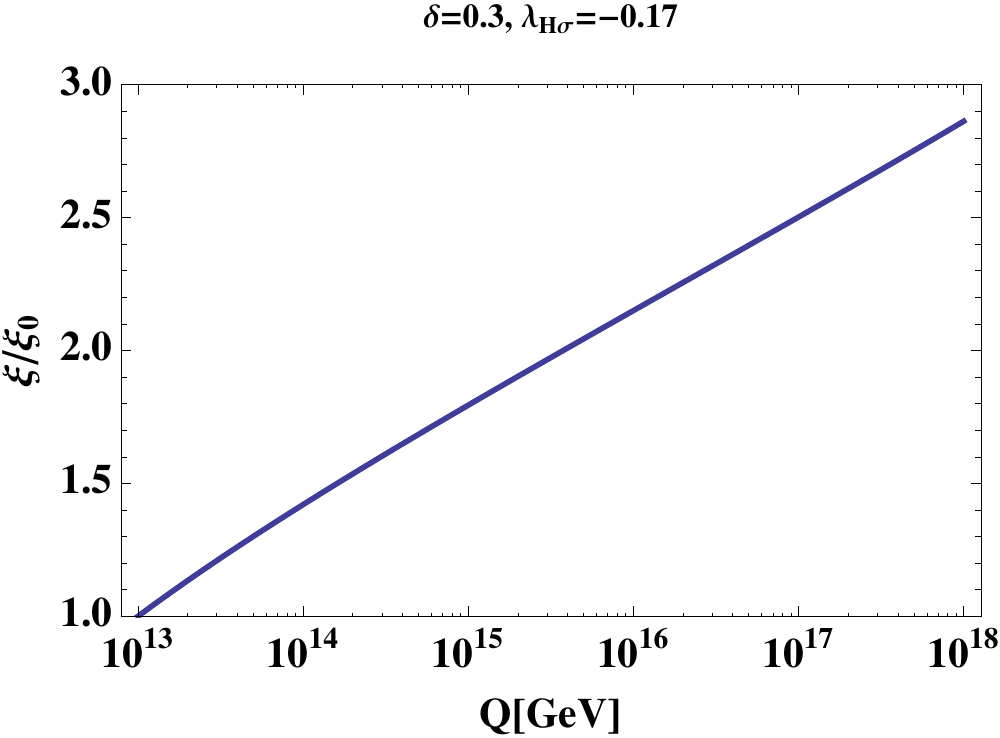}
\caption{Ratio of the loop-level to tree-level values of the non-minimal coupling consistent with the COBE normalization. We have chosen $\lambda_{\rm eff}=0.01$, $\delta=0.3$ and $\lambda_{H\sigma}=-0.17$.
}
\label{fig:xirun}
\end{figure}

On the other hand, for $N=62(59)$, the tensor-to-scalar ratio is given by
\be
r\simeq 0.0030 (0.0033).
\ee
Thus, the result is consistent with the current limit, $r<0.13$ at $95\%$ C. L., from 
WMAP $+$ eCMB $+$ BAO $+H_0$ \cite{WMAP9} and $r<0.11$ at $95\%$ C. L. from Planck data \cite{Planckdata1}.  
We note that the tensor-to-scalar ratio is insensitive to the loop corrections as the $\epsilon$ slow-roll parameter runs little as seen from eq.~(\ref{epsilon}).

Furthermore, the COBE normalization of the power spectrum constrains the inflation parameters from the following quantity evaluated at horizon exit,
\be
\Delta^2_R= \frac{25}{4} \frac{1}{24\pi^2}\Big(\frac{H^2}{{\dot\chi}}\Big)^2= \frac{25}{4} \frac{1}{72\pi^2 M^6_P} \frac{V^3}{V^{\prime 2}}.
\ee
From $\Delta^2_R=(2.464\pm 0.072)\times 10^{-9}$ \cite{WMAP9}, for the number of efoldings $N=60$, we obtain the following constraint,
\be
\xi \sqrt{\frac{\lambda_H}{\lambda_{\rm eff} \lambda_\sigma}}\simeq 47000. \label{COBE}
\ee
The running effect of the quartic couplings can be traded off for a different value of $\xi$. As shown in Fig.~\ref{fig:RG}, the effective Higgs quartic couplings can run to a larger value by order of magnitude. Then, the necessary non-minimal coupling for the COBE normalization becomes larger than what we would have obtained from the tree-level value of $\lambda_{\rm eff}$. For instance, for $\lambda_{\rm eff}=0.01$ at the sigma field threshold, the needed non-minimal coupling would be of order $10^3$ but  the RG running makes $\lambda_{\rm eff}$ order of $0.1$ so that $\xi\sim 10^4$. In Fig.~\ref{fig:xirun}, we show the ratio of loop-level to tree-level values of $\xi$ as a function of the RG scale.
From $\lambda_H=\lambda_{\rm eff}+\lambda_{H\sigma}^2/\lambda_\sigma$, we note that for $\lambda^2_{H\sigma}\gg \lambda_{\rm eff} \lambda_\sigma$, the COBE normalization condition (\ref{COBE}) with eq.~(\ref{xieff}) becomes the same as in the original Higgs inflation, 
$\xi_{\rm eff}/\sqrt{\lambda_{\rm eff}}\simeq 47000$. This is the case when the vacuum instability scale in the SM is just above the sigma scalar threshold.

\subsection{Running of spectral index}

The running of the spectral index is given by
\be
\frac{d n_s}{d\ln k}=24\epsilon^2-16 \epsilon \eta+2\zeta^2
\ee
with
\be
\zeta^2\equiv  \frac{V' V^{\prime\prime\prime}}{V^2}.
\ee
Using 
\bea
\frac{d^3 V}{d\chi^3}&=& \Big(\frac{\partial {\hat V}_0}{\partial \ln\mu} +4 {\hat V}_0\Big) \frac{\sqrt{6}}{9}
e^{-2\chi/\sqrt{6}}(1-4 e^{-2\chi/\sqrt{6}}) \nonumber \\
&&-\frac{\partial}{\partial \ln\mu} \Big(\frac{\partial {\hat V}_0}{\partial \ln\mu} +4 {\hat V}_0\Big) \,\frac{\sqrt{6}}{18}\,e^{-4\chi/\sqrt{6}}\,\left(\frac{3-4 e^{-2\chi/\sqrt{6}}}{1-e^{-2\chi/\sqrt{6}}}\right) \nonumber \\
&&+\frac{\partial^2}{\partial(\ln\mu)^2}\Big(\frac{\partial {\hat V}_0}{\partial \ln\mu} +4 {\hat V}_0\Big) \,\frac{\sqrt{6}}{36}e^{-\sqrt{6}\chi} (1-e^{-2\chi/\sqrt{6}})^{-1},
\eea
and eqs.~(\ref{epsilon}), (\ref{eta}),
and ignoring the higher derivative terms with respect to $\ln\mu$,
we obtain 
\bea
\frac{d n_s}{d\ln k}&\simeq&\frac{8\sqrt{2}}{3}\frac{1}{(2N+\sqrt{3})^2}\bigg(\frac{3}{2}-\frac{9}{2A^2}\frac{A+2}{2N+\sqrt{3}}\bigg) \nonumber \\
&&\quad+\frac{48}{(2N+\sqrt{3})^3}\bigg(\frac{2A}{|A|}+\frac{1}{A^2}\frac{6(A-1)}{2N+\sqrt{3}}\bigg) .
\eea
Consequently, for the tree-level potential with $\delta\lambda=0$, we obtain $d n_s/d \ln k=4.1\times 10^{-4}$ for $N=60$, which is consistent with the Planck$+$WMAP constraints, $d n_s/ d\ln k=-0.013\pm 0.009$ at $68\%$ C. L. within $1.5\sigma$ \cite{Planckdata1}. 
On the right of Fig.~\ref{fig:ns}, we show the dependence of the running of the spectral index on the number of efoldings and the loop corrections. We find that the loop corrections contribute to the running by $\pm\, 2\times 10^{-5}$, which is too small to be observed.

\section{Reheating temperature and number of efoldings}

From eq.~(\ref{einpot}), we rewrite the scalar potential in Einstein frame as
\be
V_E\simeq\frac{1}{4}\lambda_\sigma\omega^4 \Big(1-e^{-2\chi/\sqrt{6}M_P}\Big)^2
+\frac{1}{2}\lambda_{H\sigma} \omega^2 {\tilde\phi}^2 \Big(1-e^{-2\chi/\sqrt{6}M_P}\Big)+\frac{1}{4}\lambda_H{\tilde\phi}^4.
\ee
Here, we ignored the weak-scale Higgs mass term.
At the end of inflation, $e^{-2\chi_f/\sqrt{6}M_P}=\sqrt{3}/2$ and ${\tilde \phi}^2_f={\tilde\phi}^2_0 (1-\sqrt{3}/2)$ with ${\tilde\phi}^2_0=-\frac{\lambda_{H\sigma}}{\lambda_H}\,\omega^2$.
Therefore, both ${\tilde\phi}$ and $\chi$ carry the potential energies of order $\omega^4$ for the quartic couplings of order unity at the onset of reheating. 
By expanding the potential around $\chi=0$, we get the reheating dynamics to be a hybrid type which has both quadratic and quartic potentials with the mixing term, as follows, 
\be
V_E\simeq \frac{1}{6}\lambda_\sigma \frac{ \omega^4}{M^2_P}\chi^2+\frac{1}{4}\lambda_H{\tilde\phi}^4+\frac{1}{\sqrt{6}}\lambda_{H\sigma}\frac{\omega^2}{M_P} {\tilde\phi}^2 \chi.  \label{expanded}
\ee
The dynamics of the reheating process is much involved, so we just consider how the reheating temperature depends on the mixing coupling $\lambda_{H\sigma}$ between the Higgs and sigma fields, without going into the details.

After the biggest cosmological scale, $k^{-1}=H^{-1}_0=3000 h^{-1}\,{\rm Mpc}$, leaves the horizon, the number of efoldings is 
\be
N=\ln\Big(\frac{a_{\rm end}}{a_{\rm he}}\Big)=\ln\Big(\frac{a_{\rm end}H_{\rm end}}{a_0 H_0}\Big)-\ln \Big(\frac{H_{\rm end}}{H_{\rm he}}\Big)
\ee
where use is made of $a_{\rm he}H_{\rm he}=a_0 H_0$ in the second equality.
Assuming that slow-roll inflation is followed promptly by matter domination and consequently by radiation domination, we have 
\be
N= 62 -\ln(10^{16}\,{\rm GeV}/V^{1/4}_{\rm end})-\frac{1}{3}\ln(V^{1/4}_{\rm end}/\rho^{1/4}_{\rm reh})-\ln\Big(1-\frac{\sqrt{3}}{2}\Big)  \label{nefoldsum}
\ee
where
\bea
V_{\rm end}&=&\frac{\omega^4}{4}\Big(\lambda_\sigma-\frac{\lambda_{H\sigma}^2}{\lambda_H}\Big)\Big(1-\frac{\sqrt{3}}{2}\Big), \\
\rho_{\rm rh}&=& \frac{\pi^2 g_*}{30}\,T^4_{\rm rh}. \label{radiationrho}
\eea
Here, $g_*=106.75$ is the effective number of degrees of freedom in the SM and the inflaton vacuum energy at the end of inflation is given by $V_{\rm end}\simeq (4.8\times 10^{15}\, {\rm GeV})^4$ from the COBE normalization.

When ${\tilde\phi}_f \ll \omega$ in the end of inflation, i.e. $|\lambda_{H\sigma}|\ll \lambda_H$, the inflaton is just the sigma field with the inflaton mass being given by $m_\chi=\sqrt{(\lambda_\sigma/3)}\,M_P/\xi$. In this case, from eqs.~(\ref{einaction}) and (\ref{expanded}), the relevant interaction terms for reheating in the action in Einstein frame are
\be
{\cal L}_{\rm rh}=-\frac{\tilde\phi}{\sqrt{6}M_P}\,\partial_\mu\chi \partial^\mu {\tilde\phi}+\frac{1}{\sqrt{6}}\lambda_{H\sigma}\frac{M_P}{\xi}\chi {\tilde\phi}^2 +\frac{2}{\sqrt{6}M_P}\,\,m_{f_i} \chi {\bar f}_i f_i
\ee
where $f_i$ are the canonically normalized SM fermions. Then, the inflaton decays into a pair of $\tilde\phi$'s dominantly by both the gravitational kinetic interaction and the Higgs-portal term, with the decay rate,
\be
\Gamma(\chi\rightarrow{\tilde\phi}{\tilde\phi})=\frac{m_\chi}{192\pi}\left(\frac{m_\chi}{M_P}+2|\lambda_{H\sigma}|\frac{M_P}{\xi m_\chi}\right)^2.
\ee
So, for $2|\lambda_{H\sigma}|\sqrt{3/\lambda_\sigma}\ll m_\chi/M_P\simeq 1.2\times 10^{-5}$, where use is made of $m_\chi=2.9\times 10^{13}\,{\rm GeV}$ from the COBE normalization, the inflaton decay is dominated by the gravitational interaction. Otherwise, it is the sigma field coupling that dominantly determines the inflaton decay rate.
From $\Gamma_\chi=H_{\rm rh}=\sqrt{\rho_{\rm rh}}/(\sqrt{3}M_P)$ with eq.~(\ref{radiationrho}), the reheating temperature is given by
\be
T_{\rm rh}=\bigg(\frac{90}{\pi^2 g_*}\bigg)^{1/4}\sqrt{N_s M_P\Gamma}= \left(1+2|\lambda_{H\sigma}|\sqrt{\frac{3}{\lambda_\sigma}}\frac{M_P}{m_\chi}\right)(4.4\times 10^9\,{\rm GeV})
\ee
where the number of degrees of freedom in the Higgs doublet is taken into account by $N_s=4$.
For instance,  for $\lambda_\sigma\sim \lambda_H\sim 1$ and $|\lambda_{H\sigma}|\sim 0.01$, we would get the reheating temperature as $T_{\rm rh}\sim 10^{12}\,{\rm GeV}$.

\begin{figure}[t]
\centering
\includegraphics[width=7.5cm]{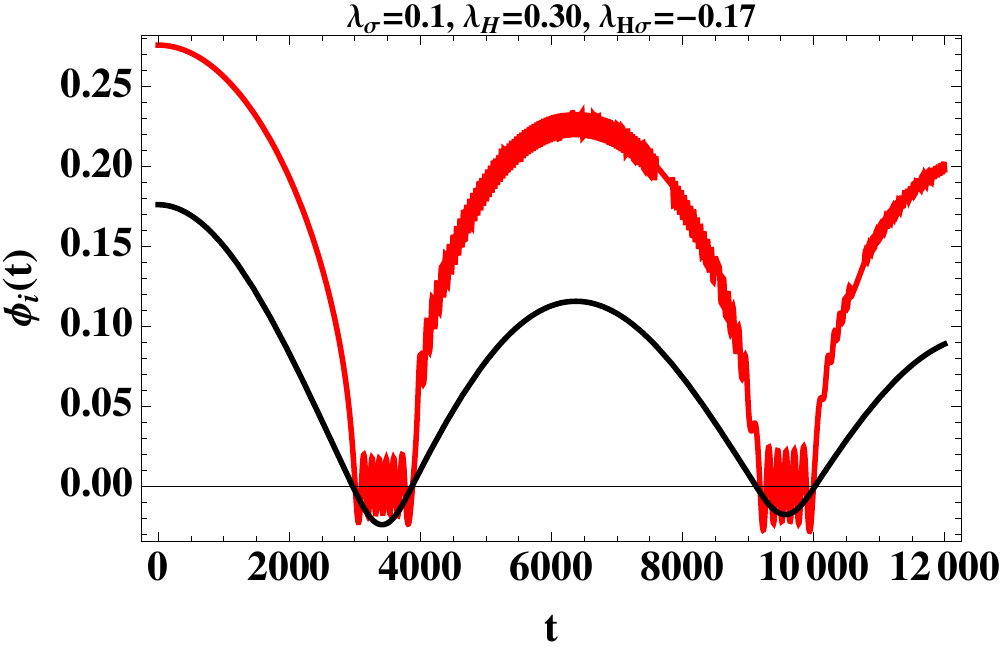}
\includegraphics[width=7.5cm]{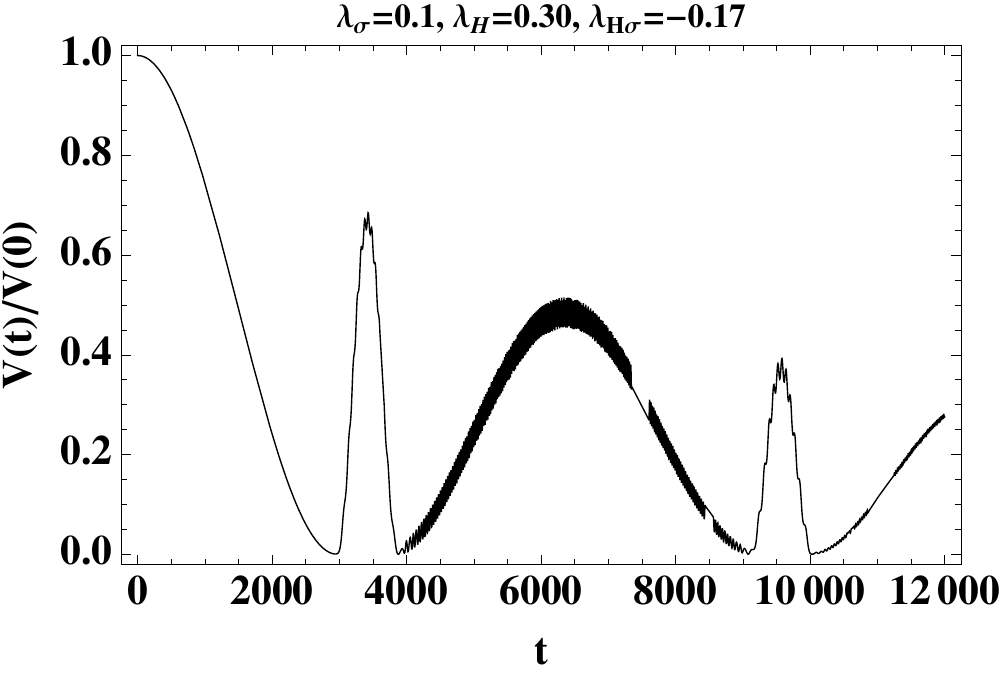}
\caption{Left: Time evolution of $\phi_i(t)=\chi(t)(\rm black),{\tilde\phi}(t)({\rm red})$. Right: Time evolution of the scalar potential $V(t)/V(0)$. In both figures, we have chosen $\lambda_\sigma=0.1, \lambda_H=0.30$ and $\lambda_{H\sigma}=-0.17$.
}
\label{fig:reheat}
\end{figure}

For a sizable $|\lambda_{H\sigma}|\sim \lambda_H$,  we get ${\tilde\phi}_f\sim \omega$, so the reheating process is similar to the SM Higgs inflation, because the Higgs carries a large fraction of the inflaton energy in the end of inflation and reheats the SM particles by the Higgs interactions. In Fig.~\ref{fig:reheat}, we show the scalar fields on the left and the scalar potential on the right as a function of time during the reheating. We note that there are multiple zero crossings of the Higgs within a single oscillation of the $\chi$ field, because $m_{\tilde\phi}\gg m_\chi$ even during the reheating.
In this case,  the SM particles can be also produced non-perturbatively in the preheating stage \cite{reheating} due to the parametric resonance in the presence of the Higgs interactions so the reheating temperature can be higher than the one obtained from the perturbative decay as in the original Higgs inflation.  The details of preheating in our model is beyond the scope of this work, so instead we quote $T_{\rm rh}= (3-15)\times 10^{13}\,{\rm GeV}$ in Higgs inflation \cite{higgsreheat} as the maximum reheating temperature possible.
Consequently, depending on the size of $\lambda_{H\sigma}$, the reheating temperature $T_{\rm rh}$ varies between $4.4\times 10^9\,{\rm GeV}$ and $1.5\times 10^{14}\,{\rm GeV}$.
Therefore, using eq.~(\ref{nefoldsum}), we obtain the number of efoldings in the following range,
\be
N=59-62.
\ee
We have used the above result as the representative values for the number of efoldings in the analysis of the previous sections.

\section{Conclusion}

We have revisited the unitarized Higgs inflation with a real scalar singlet of sigma-model type, from the perspective of the loop corrections. 
As the relevant energy scales including the inflation are much below the unitarity cutoff of the model, the Planck scale, we can use the perturbative expansion to estimate the loop corrections to the quartic couplings of the model and in turn calculate the inflationary observables. Since the mixing coupling between the sigma and Higgs fields is required to reproduce the effective large non-minimal coupling for the Higgs doublet below the sigma scalar threshold, the model has been regarded as a UV completion of the Higgs inflation in a sense of the effective action. 

The extra coupling of the new dynamical scalar contributes to the RG running of the Higgs quartic coupling, improving the vacuum stability. Furthermore, the same sigma coupling determines the RG-improved inflaton potential and controls the loop corrections to the spectral index, etc, being consistent with the nine-year WMAP data. 
There is an uncertainty in reheating temperature or the number of efoldings  in the model, because the  Higgs component of the inflaton varies depending on the sigma coupling.  We conclude that the loop corrections to the spectral index in the model are under control within the uncertainties in the reheating temperature.

\section*{Acknowledgments}
The author thanks Fedor Bezrukov for the early discussion on the loop corrections and the reheating dynamics in Higgs inflation and Wan-Il Park for useful comments.

\def\theequation{A.\arabic{equation}}

\setcounter{equation}{0}

\vskip0.8cm
\noindent
{\Large \bf Appendix A:  Renormalization group equations}
\vskip0.4cm
\noindent

We take into account the effects of a non-minimal coupling to gravity of the sigma and Higgs fields through suppression factors in the RGE \cite{RG2}.  The one-loop RG evolution of the scalar quartic couplings above the sigma-field threshold is governed by
\bea
(4\pi)^2\frac{d\lambda _{{H}}}{d\ln\mu} &=& (12 y_t^2-3 {g'}^2-9 g^2) \lambda _{{H}}-6 y_t^4+\frac{3}{8}\Big(2g^4+ ({g'}^2+g^2)^2\Big)  \nonumber \\
&&+(18c^2_h+6) \lambda _{{H}}^2+2c^2_\sigma \lambda _{{H\sigma}}^2, \label{lambda1}   \\
(4\pi)^2\frac{d\lambda _{{H\sigma}}}{d\ln\mu} &=& \frac{1}{2}\lambda _{{H\sigma}} (12 y_t^2-3g^{\prime 2}-9g^2+12(1+c^2_h) \lambda _{{H}}+12c^2_\sigma \lambda _{{\sigma}})+8 c_h c_\sigma\lambda _{{H\sigma}}^2,  \label{lambda2} \\
(4\pi)^2\frac{d\lambda _{{\sigma}}}{d\ln\mu} &=& 2(3+c^2_h) \lambda _{{H\sigma}}^2+18 c^2_\sigma\lambda _{{\sigma}}^2.  \label{lambda3}
\eea
The two-loop RG equations for the gauge and Yukawa couplings are
\bea
(4\pi)^2 \frac{d g'}{d\ln\mu}&=&\frac{81+c_h}{12} g^{' 3}+\frac{g^{'3}}{16\pi^2} \Big(\frac{199g^{'2}}{18}+\frac{9g^2}{2}+\frac{44 g^2_3}{3}-\frac{17 c_h y^2_t}{6}\Big), \\
(4\pi)^2 \frac{d g}{d\ln\mu}&=&-\frac{39-c_h}{12} g^3+\frac{g^3}{16\pi^2}\Big(\frac{3}{2}g^{'2}+\frac{35}{6}g^2+12 g^2_3-\frac{3}{2} c_h y^2_t\Big), \\
(4\pi)^2 \frac{d g_3}{d\ln\mu}&=& -7 g^3_3+
\frac{g^3}{16\pi^2}\Big(\frac{11}{6}g^{'2}+\frac{9}{2} g^2-26 g^2_3-2 c_h y^2_t\Big), 
\eea
and
\bea
(4\pi)^2 \frac{d y_t}{d\ln\mu}&=& y_t \Big(\Big(\frac{23}{6}+\frac{2}{3} c_h\Big)y_t^2- 8 g^2_3 - \frac{9}{4}g^2-\frac{17}{12} g^{'2}\Big) \nonumber \\
&&+\frac{y_t}{16\pi^2}\bigg[ -\frac{23}{4} g^4-\frac{3}{4}g^2 g^{'2}+\frac{1187}{216}g^{'4}+9 g^2 g^2_3 \nonumber \\
&&\quad+\frac{19}{9}g^{\prime 2}g^2_3 -108 g^4_3+c_h y^2_t\Big(\frac{225}{16}g^2+\frac{131}{16} g^{\prime 2} +36 g^2_3\Big) \nonumber \\
&&\quad + 6(-2c^2_h y^4_t -2 c^3_h y^2_t\lambda_H + c^2_h \lambda^2_H)  \bigg]. 
\eea

The RG equations for non-minimal couplings are
\bea
(4\pi)^2 \frac{d\xi}{d\ln\mu}&=& 2(3+c_h)\lambda_{H\sigma}\Big(\zeta+\frac{1}{6}\Big) +6c_\sigma  \lambda_\sigma\Big(\xi+\frac{1}{6}\Big), \\
(4\pi)^2 \frac{d\zeta}{d\ln\mu}&=& \Big((6+6c_h)\lambda_H+6y^2_t  -\frac{3}{2}(3g^2+g^{\prime 2})\Big) \Big(\zeta+\frac{1}{6}\Big)+2c_\sigma \lambda_{H\sigma} \Big(\xi+\frac{1}{6}\Big).
\eea

The suppression factors, $c_\sigma$ and $c_h$, are given in terms of the Weyl rescaling factor 
$\Omega^2=\xi\sigma^2/M^2_P$ as
$c_\sigma= \Omega^{-2} (\partial\chi/\partial \sigma)^{-2}=1/(6\xi)\ll 1$ and
$c_h= \Omega^{-2} (\partial\phi/\partial h)^{-2}=1$, where $\chi$ and $\phi$ are canonical fields.
Therefore, the running of $\lambda_H$ is SM-like, as loops containing the sigma scalar are suppressed for $c_\sigma \ll 1$ and Higgs loops are the same as in the SM. 
The suppressed sigma-field loops can help to keep the non-minimal coupling $\zeta$ of the Higgs  small under the RG evolution.

Using the RG equations (\ref{lambda1})-(\ref{lambda3}), we also get the RG equation for $\lambda_{\rm eff}\equiv \lambda_H-\lambda^2_{H\sigma}/\lambda_\sigma$, with $\delta\lambda \equiv \lambda^2_{H\sigma}/\lambda_\sigma$, as
\bea
\frac{d\lambda_{\rm eff} }{d\ln\mu} \approx \beta^{\rm SM}_{\lambda_{\rm eff}}
+\frac{8}{(4\pi)^2}(3\lambda_{\rm eff}+\delta\lambda)\,\delta\lambda\,.
\eea

\end{document}